\documentclass[a4paper,twocolumn,fleqn,superscriptaddress,preprintnumbers,prd]{revtex4}

\usepackage{graphicx,color}
\usepackage{amsmath,amssymb,amsfonts}
\usepackage{stackrel}

\providecommand{\openone}{\leavevmode\hbox{\large1\kern-7.3pt\normalsize1}}
\newcommand{\be}{\begin{equation}}
\newcommand{\ee}{\end{equation}}
\newcommand{\ba}{\begin{eqnarray}}
\newcommand{\ea}{\end{eqnarray}}
\newcommand{\rmi}[1]{{\mbox{\scriptsize #1}}}

\newcommand{\tr}{{\rm Tr\,}}
\newcommand{\nn}{\nonumber \\}
\newcommand{\fr}[2]{{\frac{#1}{#2}\,}}
\newcommand{\msbar}{{\overline{\mbox{\rm MS}}}}

\renewcommand{\(}{\left(}
\renewcommand{\)}{\right)}

\newcommand{\e}{\epsilon}

\def\sumint{\hbox{$\sum$}\!\!\!\!\!\!\!\int}

\renewcommand{\ln}{{\rm ln}}
\newcommand{\mubar}{\bar{\mu}}

\newcommand{\imathb}{i}

\newcommand{\im}{{\rm Im}\,}

\begin{document}

\title{Cool quark matter}

\preprint{CERN-TH/2016-045, HIP-2016-06/TH}
\author{Aleksi Kurkela}
\affiliation{Theoretical Physics Department, CERN, Geneva, Switzerland and \\ Faculty of Science and Technology, University of Stavanger, 4036 Stavanger, Norway}
\author{Aleksi Vuorinen}
\affiliation{Department of Physics and Helsinki Institute of Physics, P.O.~Box 64, FI-00014 University of Helsinki, Finland}

\begin{abstract}

\noindent We generalize the state-of-the-art perturbative Equation of State of cold quark matter to nonzero temperatures, needed in the description of neutron star mergers and core collapse processes. The new result is accurate to ${\mathcal O}(g^5)$ in the gauge coupling, and is based on a novel framework for dealing with the infrared sensitive soft field modes of the theory. The zero Matsubara mode sector is treated via a dimensionally reduced effective theory, while the soft non-zero modes are resummed using the Hard Thermal Loop approximation. This combination of known effective descriptions offers unprecedented access to small but nonzero temperatures, both in and out of beta equilibrium.

\end{abstract}

\maketitle

\section{Introduction}

The recent discovery of gravitational waves emitted by two merging black holes by the LIGO and Virgo collaborations has opened up a new observational window in astrophysics \cite{Abbott:2016blz}. It is expected that in the near future, a similar signal will be detected from the merger of two neutron stars or a neutron star and a black hole, or from a supernova explosion. This would lead to a wealth of new information about the properties of neutron stars and the matter they are composed of \cite{Andersson:2009yt}, highlighting the need to understand the material properties of dense nuclear matter from its microscopic description.

Figuring out the properties of dense nuclear and quark matter is a notoriously difficult task, as it necessitates a nonperturbative treatment of the theory of strong interactions, QCD, at large baryon chemical potentials $\mu_\rmi{B}$ \cite{Brambilla:2014jmp}. At the moment, the Equation of State (EoS) of zero-temperature nuclear matter is under control up to roughly the nuclear saturation density $n_s\approx 0.16/\text{fm}^3$ \cite{Tews:2012fj}, beyond which it is typically approximated by a polytropic EoS \cite{Hebeler:2013nza}. As recently demonstrated \cite{Kurkela:2014vha,Fraga:2015xha}, the properties of these polytropes can furthermore be significantly constrained using the perturbative EoS of zero-temperature quark matter \cite{Kurkela:2009gj}, known up to order $g^4 = (4\pi \alpha_s)^2$ in the strong coupling constant (see also \cite{fmcl,avpres,Fraga:2001id}).

For quiescent neutron stars, the approximation of working at exactly zero temperature is typically rather good. In the description of violent phenomena, such as neutron star mergers, thermal corrections to the EoS are, however, absolutely essential to include, as temperatures up to ca.~100 MeV may occur \cite{Shen:1998gq}. It therefore becomes necessary to also account for finite-$T$ effects in the properties of quark matter using perturbation theory --- a task complicated by nonlinear infrared (IR) dynamics. 

The reason for the appearance of IR problems in perturbative calculations lies in the medium modifications that long wavelength excitations receive. In order to identify the modes needing nonperturbative treatment, consider the dispersion relation of gluon fields \footnote{The fermionic case proceeds in a qualitatively similar way. It, however, turns out that to the order we are working in, no resummation is required.}, which has the parametric form $-\omega^2 + k^2 + \Pi(\omega, k)=0$, with $\Pi$ representing a given component of the one-loop polarization tensor. This quantity has the parametric order of the in-medium screening mass
\ba
m_\rmi{E}^2&=& \fr{g^2}{3} \Bigg[\Big(N_c+\frac{N_f}{2}\Big) T^2 + \fr{3}{2\pi^2}\!\sum_f \mu_f^2 \Bigg], \label{mEres}
\ea
where $\mu_f$ stand for the quark chemical potentials. For the majority of modes, $k\gg m_E$, and medium modifications represent only a small perturbation to the dispersion relation, implying that a ``naive'' weak coupling (loop) expansion in $g^2$ can be carried out. However, when the medium modification becomes an $\mathcal{O}(1)$ effect, i.e., $-\omega^2 + k^2 \lesssim m_E^2$, it must be treated nonperturbatively.

In the evaluation of bulk thermodynamic quantities, $\omega$ takes values at imaginary Matsubara frequencies $i\omega_n$, with $\omega_n=2\pi n T$ for bosons and $(2n + 1)\pi T -i\mu_f$ for fermions. For $T\gg m_\rmi{E}$, it is only the bosonic $n=0$ mode that must be treated nonperturbatively, using either the dimensionally reduced (DR) effective theory Electrostatic QCD (EQCD) \cite{Appelquist:1981vg,Kajantie:1995dw,Braaten:1995cm} or the Hard Thermal Loop (HTL) framework \cite{Braaten:1989mz,Braaten:1991gm}. This has led to an ${\mathcal O}(g^6\,\ln\,g)$ result for the high-temperature EoS \cite{klry,avpres}, as well as a significant improvement of the convergence of the weak coupling expansion
\cite{Blaizot:2000fc,Blaizot:2003iq,Laine:2006cp,Andersen:2011sf,Haque:2014rua,Mogliacci:2013mca}. At lower temperatures, in particular when $T$ becomes of order $m_\rmi{E}\sim g\mu_\rmi{B}$, an increasing set of low-lying Matsubara modes, however, needs to be resummed. This poses a problem, which has been tackled in the regime $T\sim g^x\mu_\rmi{B}$, $x>1$, by the Hard Dense Loop (HDL) approach, revealing non-Fermi liquid behavior \cite{Ipp:2003cj,Schafer:2004zf,Gerhold:2004tb,Gerhold:2005uu}.

At present, the only ${\mathcal O}(g^4)$ result available for the EoS at all temperatures is based on a tour-de-force resummation that applies the one-loop gluon polarization tensor of the full theory \cite{Ipp:2006ij}. This calculation made no use of the fact that even at low temperatures only soft gluon modes that need to be resummed, or that the self-energies obtain their dominant contributions from the hard scale, i.e.~from HTL kinematics. This resulted in a cumbersome numerical result, only worked out for three massless quark flavors at equal chemical potentials.

In this letter, we make use of the two effective descriptions for the soft sector of QCD mentioned above --- EQCD and Hard Thermal Loops --- to formulate a simple framework for determining bulk thermodynamic quantities at all values of $T/\mu_\rmi{B}$. In particular, this development improves the current situation in the region of small but nonzero temperatures, which now becomes smoothly connected to the limits of $T=0$ and $T\gtrsim \mu_B$. 

\section{Methodology}

Consider the weak coupling expansion of the QCD pressure as a function of the temperature $T$ and the quark chemical potentials $\mu_f$. Denoting by $p_\rmi{QCD}^\rmi{res}$ an expression for the quantity, where sufficient resummations have been carried out so that the result contains all contributions up to the desired order in $g$, we may add and subtract from it a function $p_\rmi{soft}^\rmi{res}$. This term is defined as the resummed contribution of all soft modes requiring nonperturbative treatment, such that the difference $p_\rmi{QCD}^\rmi{res}-p_\rmi{soft}^\rmi{res}$ only contains contributions from hard modes. This implies that we may evaluate both terms in the difference in a naive loop expansion \cite{Laine:2011xm}, giving
\ba
p_\rmi{QCD}^\rmi{res} = p_\rmi{QCD}^\rmi{res} - p_\rmi{soft}^\rmi{res} + p_\rmi{soft}^\rmi{res} 
=  p_\rmi{QCD}^\rmi{naive} - p_\rmi{soft}^\rmi{naive} + p_\rmi{soft}^\rmi{res}.\hspace{0.2cm}  \label{starting}
\ea
Despite its trivial appearance, this relation contains a remarkable simplification, as it expresses the contribution of the hard modes through a loop expansion, available in the literature \cite{avpres}. This reduces the problem of evaluating the EoS to properly identifyig the soft sector, as well determining the functions $p_\rmi{soft}^\rmi{res}$ and $p_\rmi{soft}^\rmi{naive}$.

A useful feature of the above formulation is that eq.~(\ref{starting}) is insensitive to the exact definition of the ``soft'' sector as long as it contains all the modes that need to be resummed. Should some hard contributions be included in $p_\rmi{soft}^\rmi{res}$, they get subtracted by $p_\rmi{soft}^\rmi{naive}$, removing any possible overcountings. A minimal description of the soft physics, applicable at all temperatures and densities, is to handle the static ($n=0$ bosonic) sector via the dimensionally reduced effective theory EQCD \cite{Braaten:1995cm}, while treating the non-static modes with $k\sim m_\rmi{E}$ using an HTL expansion \cite{Andersen:1999sf}. This allows us to write eq.~(\ref{starting}) in the form
\ba
\!\!\!\! p_\rmi{QCD} &=& p_\rmi{QCD}^\rmi{naive} + \underbrace{p_\rmi{DR}^\rmi{res} - p_\rmi{DR}^\rmi{naive}}_{p_\rmi{DR}^\rmi{corr}} + \underbrace{p_\rmi{HTL}^\rmi{res} -  p_\rmi{HTL}^\rmi{naive}}_{p_\rmi{HTL}^\rmi{corr}}, \label{resgen}
\ea
where it is understood that the HTL formulation is only used for the non-static modes. We have also defined two UV finite functions, $p_\rmi{DR}^\rmi{corr}$ and $p_\rmi{HTL}^\rmi{corr}$, which will turn out very convenient for our discussion. In accordance with \cite{Ipp:2006ij}, we shall observe that the HTL sector only contributes in the regime of low temperatures, $T\lesssim m_\rmi{E}$, and that the DR resummation alone suffices for larger values of $T$. In the following, we briefly discuss the three parts of eq.~(\ref{resgen}).

\textit{The naive QCD pressure.} As noted above, the term $p_\rmi{QCD}^\rmi{naive}$ is obtainable through a strict loop expansion of the pressure within the full theory. Its definition thereby coincides with that of the parameter $p_\rmi{E}$ of EQCD \cite{Braaten:1995cm}, which has been determined up to three-loop, or $g^4$, order at all $T$ and $\mu$ in \cite{avpres,avthesis}, utilizing techniques developed in \cite{az}. The result can be directly read off from eqs.~(3.6)--(3.14) of \cite{avpres}, in which a typo was later spotted and corrected in \cite{Haque:2013sja}. The somewhat lengthy expressions for $p_\rmi{QCD}^\rmi{naive}$ and its low-temperature limit are reproduced in Appendix B of this article (see also Appendix A for our notation).

It should be noted that $p_\rmi{QCD}^\rmi{naive}$ contains in principle both UV and IR divergences. The UV poles are removed by renormalization and are not visible in the result. The IR divergences are on the other hand physical, and cancel against equal but opposite ones in $p_\rmi{DR}^\rmi{corr}$ and $p_\rmi{HTL}^\rmi{corr}$. The IR divergences that cancel against those of $p_\rmi{DR}^\rmi{corr}$ are the $1/\epsilon$ terms on the first two lines of eq.~\ref{ae3} (we work in the $\msbar$ scheme in $d=3-2\epsilon$ spatial dimensions, applying dimensional regularization). At the same time, the IR sensitivity of  $p_\rmi{QCD}^\rmi{naive}$ that cancels against $p_\rmi{HTL}^\rmi{corr}$ is manifested in the $\ln\,T$ term in eq.~(\ref{p3smallT}), diverging as $T\to 0$. 

\textit{The dimensionally reduced term.} The function $p_\rmi{DR}^\rmi{res}$ denotes the contribution of the Matsubara zero mode sector to the pressure, and can be evaluated using a combination of a weak coupling expansion within the effective theory EQCD as well as three-dimensional lattice simulations that become necessary at order $g^6$ \cite{Kajantie:2003ax,Hietanen:2006rc,DiRenzo:2006nh}. For consistency, we shall only quote the (analytically known) result to ${\mathcal O}(g^5)$ here, as other contributions of ${\mathcal O}(g^6)$ are in any case missing from our result. This produces
\ba
p_\rmi{DR}^\rmi{res}/T&=&\frac{d_A}{12\pi}m_\rmi{E}^3 \label{pDR}\\
&+&\frac{d_A C_A}{(4\pi)^2} g_\rmi{E}^2 m_\rmi{E}^2\bigg[-\frac{1}{4\epsilon}-\frac{3}{4}-\ln\,\frac{\bar{\Lambda}}{2m_\rmi{E}}\bigg] \nn
&+&\frac{d_A C_A^2}{(4\pi)^3} g_\rmi{E}^4m_\rmi{E}\bigg[-\frac{89}{24}-\frac{\pi^2}{6}+\frac{11}{6}\ln\,2\bigg], \nonumber
\ea
where $\bar{\Lambda}$ is the renormalization scale and $d_A\equiv N_c^2-1$, $C_A\equiv N_c$. The leading-order result for $m_\rmi{E}$ is given in eq.~(\ref{mEres}) above, while the EQCD gauge coupling $g_\rmi{E}$ has the form $g_\rmi{E}^2=g^2T +{\mathcal O}(g^4)$. Higher order corrections to these parameters are available in \cite{avpres}.

The $1/\epsilon$ pole on the second line of eq.~(\ref{pDR}) is of UV nature.  It coincides with the UV divergence of the unresummed function $p_\rmi{DR}^\rmi{naive}$, thus making the combination $p_\rmi{DR}^\rmi{corr}$ UV safe. It, however, turns out that $p_\rmi{DR}^\rmi{naive}$ also contains an equal but opposite IR divergence (identifiable with that of $p_\rmi{QCD}^\rmi{naive}$) and moreover completely vanishes in dimensional regularization where the same parameter $\epsilon$ is used to regulate both UV and IR divergences \footnote{In the evaluation of the quantity $p_\rmi{DR}^\rmi{naive}$, the mass term of the $A_0$ field in EQCD is treated as a perturbation. This leads to the pressure being expressible in terms of scalefree integrals of the type $\int_p \frac{1}{(p^2)^n}$, which all vanish in dimensional regularization.}. To this end, we are left with the identity $p_\rmi{DR}^\rmi{corr}=p_\rmi{DR}^\rmi{res}$, where the $1/\epsilon$ divergence of the function is now identified as an IR pole.

\begin{figure}[t]
\begin{center}
\includegraphics[width=0.38\textwidth]{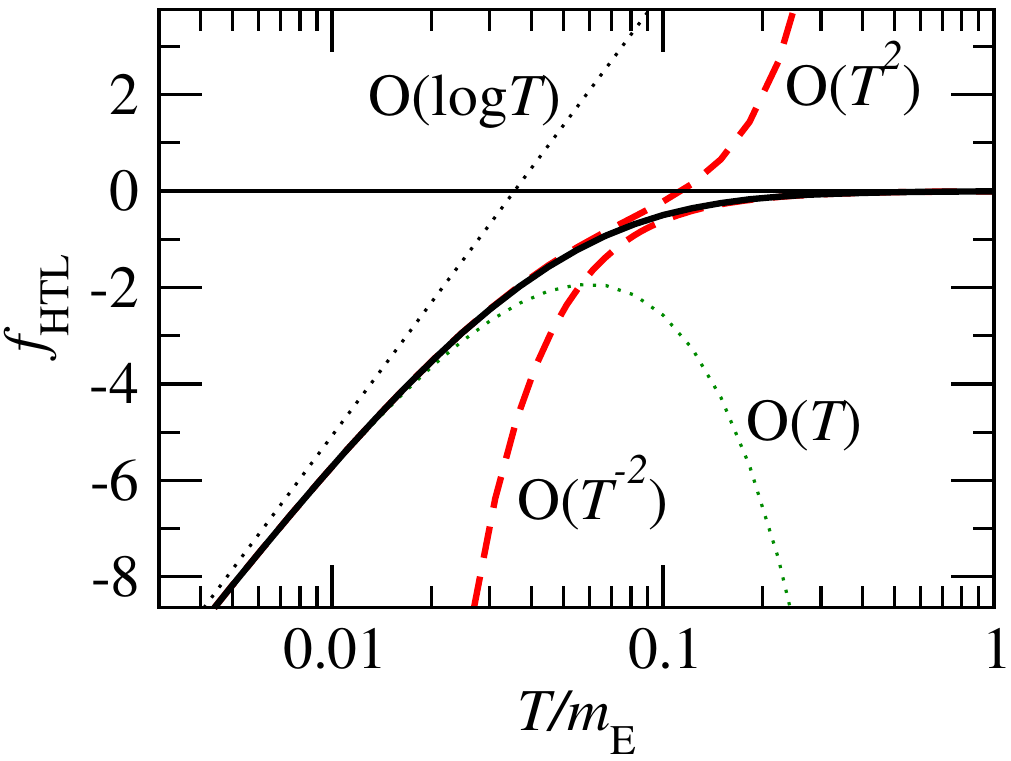}
\caption{The behavior of the function $f_\rmi{HTL}(x)$, defined in eq.~(\ref{pHTLres}). Shown here are also the first three orders of the small-$T$ expansion as well as the leading high-$T$ limit, as indicated by eqs.~(\ref{smallx}) and (\ref{largex}).}\label{fig:1}
\end{center}
\end{figure}

\textit{The HTL contributions.} The resummed HTL contribution to the pressure takes the form of the familiar ``HTL ring sum'' integral \cite{Andersen:1999sf}
\ba
p_\rmi{HTL}^\rmi{res}&=&  -\frac{(d-1)d_A}{2} \sumint^{'}_K  \log \left[ 1 + \frac{\Pi_\rmi{T}(K)}{K^2} \right] \nn 
&-& \frac{d_A}{2} \sumint^{'}_K  \log \left[ 1 + \frac{\Pi_\rmi{L}(K)}{K^2} \right], \label{psoftn}
\ea
where the primes remind us of the fact that the zero mode is to be left out from the corresponding Matsubara sums. The functions $\Pi_\rmi{T/L}$ stand here for the transverse and longitudinal HTL self-energies
\begin{align}
\frac{\Pi_\rmi{T}(K)}{K^2} &= \frac{m_\infty^2}{K^2} - \frac{1}{2}\Pi_\rmi{HTL}(K), \label{pitdef} \\
\frac{\Pi_\rmi{L}(K)}{K^2} &=  \Pi_\rmi{HTL}(K), \label{pildef}
\end{align}
with $m_\infty^2\equiv m_\rmi{E}^2/(d-1)$ and (in exactly three dimensions)
\ba
 \!\!\!\!\!\!\!\!\!\!\Pi_\rmi{HTL}(\omega,k)  &=& m_\rmi{E}^2\,\Bigg[ \frac{1}{k^2}-\frac{\omega}{2 k^3} \log\left[\frac{\omega+i 0^+ \! +k}{\omega+i 0^+ \! -k}\right]\Bigg].
\ea
The corresponding naive HTL contribution is on the other hand obtained by simply expanding the logarithms of eq.~(\ref{psoftn}) in powers of the self-energies, which produces
\ba
p_\rmi{HTL}^\rmi{naive}&=& -d_A \sumint'_k\bigg[  \frac{d-1}{2}\frac{ \Pi_\rmi{T}}{K^2} + \frac{1}{2}\frac{\Pi_\rmi{L}}{K^2}\label{psoftn2} \\
&-&\frac{1}{2}\left( \frac{d-1}{2}\frac{\Pi_\rmi{T}^2}{(K^2)^2} +\frac{1}{2}\frac{\Pi_\rmi{L}^2}{(K^2)^2} \right)\bigg] +{\mathcal O}(g^6). \nonumber
\ea

The functions $p_\rmi{HTL}^\rmi{res}$ and $p_\rmi{HTL}^\rmi{naive}$ are clearly both IR finite at nonzero $T$, but contain UV divergences that however cancel in the combination $p_\rmi{HTL}^\rmi{corr}$ defined in eq.~(\ref{resgen}). The numerical evaluation of this function follows the treatment of \cite{Andersen:1999sf} and is briefly discussed in Appendix C below. The result takes the form
\ba
p_\rmi{HTL}^\rmi{corr}&=&\frac{d_A m_\rmi{E}^4}{256\pi^2}\, f_\rmi{HTL}(T/m_\rmi{E}),\label{pHTLres}
\ea
where the numerically determined function $f_\rmi{HTL}$, displayed in fig.~1, has the limiting values 
\ba
f_{\rmi{HTL}}(x)&\xrightarrow[x\to 0]{}&4\,\ln\,x+11-4\gamma-\frac{2\pi^2}{3}+\frac{14\,\ln\,2}{3} \nn
&&+\frac{16\,\ln^2\,2}{3}+4\,\ln\,\pi-\delta-\frac{64\pi}{3}x\nn
&&-\frac{32\pi^2}{9}x^2\bigg(\,\ln\,x-\ln\,\frac{4}{\pi}-\gamma+\frac{\zeta'(2)}{\zeta(2)}\bigg)\nn
&&+{\mathcal O}(x^{8/3}), \label{smallx} \\
f_{\rmi{HTL}}(x)&\xrightarrow[x\to \infty]{}&-\frac{0.006178(1)}{x^2}+{\mathcal O}(1/x^{3}),\label{largex}
\ea
with $\delta\approx -0.8563832$ \cite{avpres}. Some higher order terms to the former  expansion can be obtained from \cite{Gerhold:2004tb,Gerhold:2005uu}.

\begin{figure}[t]
\begin{center}
\includegraphics[width=0.38\textwidth]{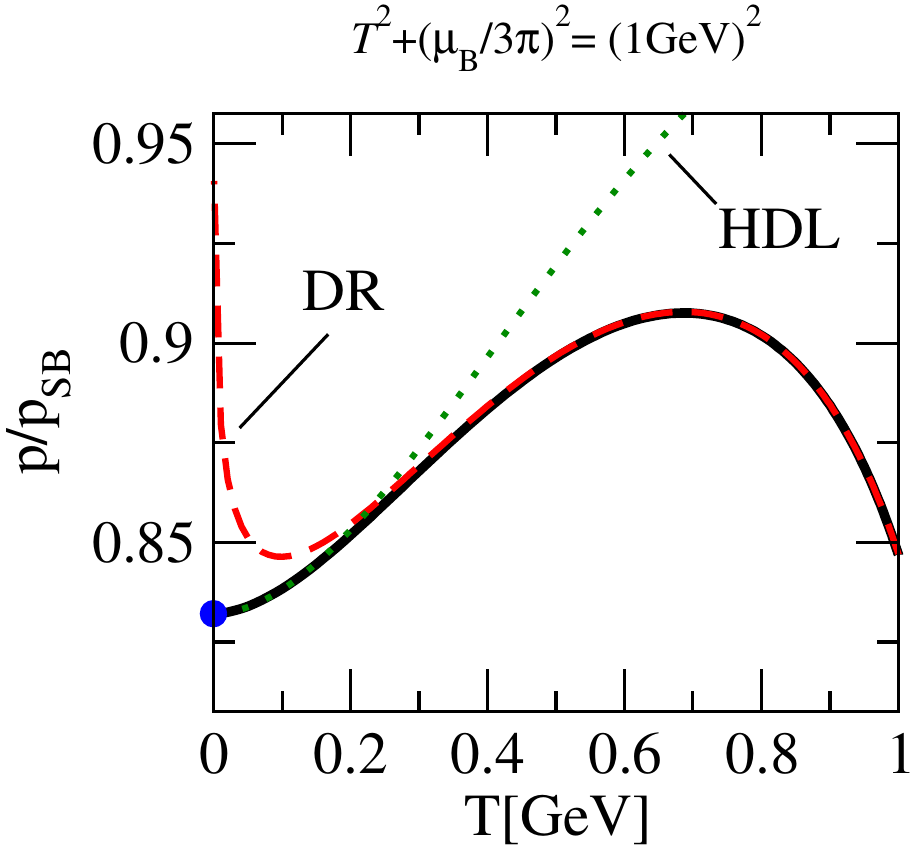}
\caption{The pressure of QCD normalized by its free 
(Stefan-Boltzmann) limit, evaluated for a fixed value of the function $T^2+(\mu_\rmi{B}/3\pi)^2$.  The black curve corresponds to our new result, while the red dashed line stands for the DR prediction of \cite{avpres}, the green dotted line for the HDL result of \cite{Gerhold:2005uu}, and the single blue dot for the $T=0$ limit of \cite{fmcl,avpres}. The scale $\bar{\Lambda}$ is set to its midpoint value here, specified in the main text.}\label{fig:2}
\end{center}
\end{figure}

\section{Results}

\begin{figure*}[t]
\begin{center}
\includegraphics[width=1.0\textwidth]{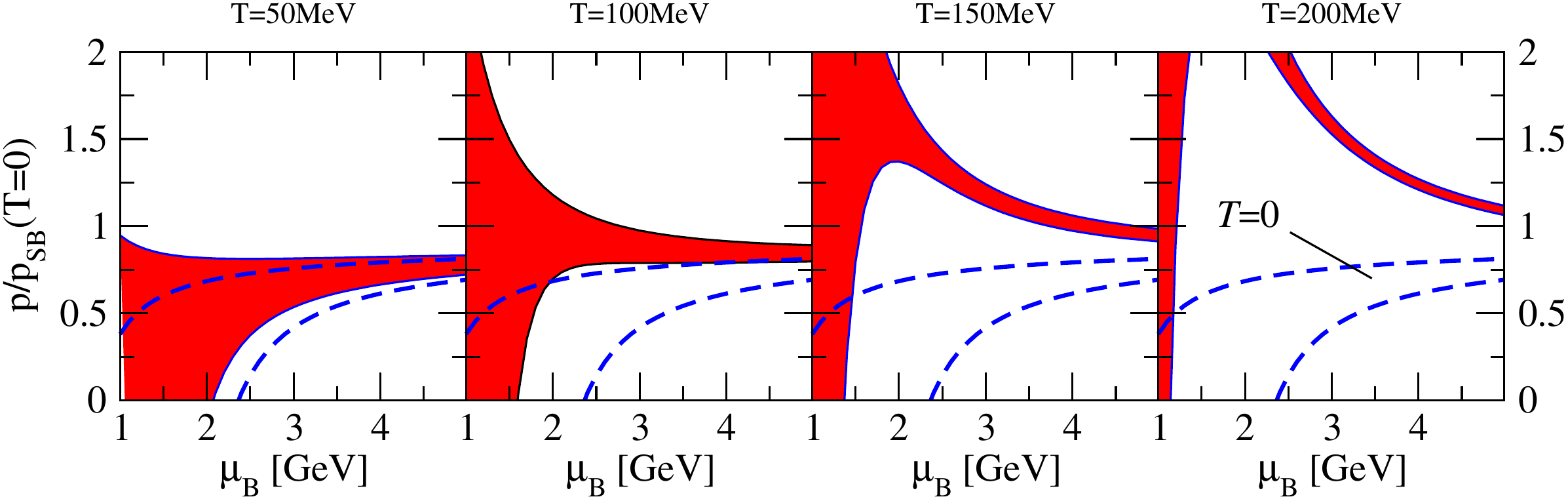}
\caption{The pressure of deconfined quark matter given as a function of $\mu_\rmi{B}$ for four different temperatures, and normalized to the pressure of a system of free quarks at $T=0$. The red bands correspond to our new result, eq.~(\ref{resgen}), with their widths originating from a variation of the renormalization scale $\bar{\Lambda}$ as explained in the main text. The dashed blue lines indicate the corresponding  ${\mathcal O}(g^4)$ result at zero temperature \cite{fmcl,avpres}.} \label{fig:3}
\end{center}
\end{figure*}

At this point, we have evaluated all three parts of $p_\rmi{QCD}$ in eq.~(\ref{resgen}), i.e.~$p_\rmi{QCD}^\rmi{naive}$, $p_\rmi{DR}^\rmi{corr}=p_\rmi{DR}^\rmi{res}$, and $p_\rmi{HTL}^\rmi{corr}$. Below, we briefly discuss the structure of this combination in two different regimes: $T\gg m_\rmi{E}$ and $T\lesssim m_\rmi{E}$, or high and low temperatures, respectively.

\textit{High temperatures}: When $T$ is parametrically larger than $m_\rmi{E}$, in particular of ${\mathcal O}(\mu_\rmi{B})$, we see from eqs.~(\ref{pHTLres}) and (\ref{largex}) that the HTL contribution to the pressure becomes of ${\mathcal O}(g^6)$ and is thus no longer interesting for us. This is a manifestation of the fact that the HTL resummation was only carried out for the non-zero Matsubara frequencies, which all become \textit{hard} modes at high $T$. Recalling further that we may associate $p_\rmi{QCD}^\rmi{naive}$ and $p_\rmi{DR}^\rmi{res}$ with the functions $p_\rmi{E}$ and $p_\rmi{M}$ of EQCD, we see that our result exactly reduces to the known high-temperature one of \cite{avpres}, worked out up to and including ${\mathcal O}(g^6\ln\,g)$ there.

\textit{Low temperatures}: Proceeding to the opposite $T\to 0$ limit, the naive QCD contribution to the pressure reduces to eqs.~(\ref{p1smallT})--(\ref{p3smallT}), while the three terms of $p_\rmi{DR}^\rmi{res}$, visible in eq.~(\ref{pDR}), are suppressed by factors of ${\mathcal O}(T)$, ${\mathcal O}(T^2)$ and ${\mathcal O}(T^3)$, respectively. Adding to this the first orders of the low-temperature expansion of the HTL contribution, eq.~(\ref{smallx}), we witness the cancelation of the $\ln\,T$ terms of $p_\rmi{QCD}^\rmi{naive}$ and $p_\rmi{HTL}^\rmi{corr}$, while the other terms surviving in the $T=0$ limit exactly reproduce the well-known result of \cite{fmcl,avpres}. The leading correction to this expression turns out not to be of linear order in $T$, as the ${\mathcal O}(T)$ contributions to $p_\rmi{DR}^\rmi{res}$ and $p_\rmi{HTL}^\rmi{corr}$ cancel each other, but the lowest nonvanishing corrections are of ${\mathcal O}(T^2\ln\,T)$. These logarithmic terms have been thoroughly analyzed in \cite{Gerhold:2004tb,Gerhold:2005uu}. 
Interestingly, at higher orders in the expansion of the low-$T$ pressure, the ${\mathcal O}(g^4)$ correction to $m_\rmi{E}^2$ produces a contribution of order $g^5 T\,\ln\,T$ through the first term of eq.~(\ref{pDR}). We expect, however, that this ${\mathcal O}(g^6)$ term gets canceled by a similar correction to the HTL term of eq.~(\ref{pHTLres}).

A crucial feature of our new EoS is that due to its simple form, it is immediately amenable for numerical evaluation, as well as for a resummation along the lines of refs.~\cite{Blaizot:2003iq,Laine:2006cp}. Studying first the generic form of the pressure for $N_c=N_f=3$, we display in fig.~\ref{fig:2} the smooth interpolation of our result between the low-temperature HDL-approximation of \cite{Gerhold:2005uu} and the high-temperature EQCD result of \cite{avpres}, when the Root Mean Square (RMS) of the scales $T$ and $\mu_\rmi{B}/(3\pi)$ is set to 1 GeV, and the temperature is increased.

In fig.~\ref{fig:3}, we next look at the form of our result for fixed values of the temperature, $T=50$, 100, 150 and 200 MeV. Shown here are also the effects of varying the $\msbar$ renormalization scale $\bar{\Lambda}$ by a factor of 2 around the RMS of the commonly used $\mu_B=0$ and $T=0$ scales $\bar{\Lambda}=0.723\times 4\pi T$ \cite{klry} and $\bar{\Lambda}=2\mu_\rmi{B}/3$ \cite{Kurkela:2009gj}. Just like in fig.~2, we have applied here the two-loop running coupling and the value 378 MeV for $\Lambda_\rmi{QCD}$. The fast increase of the uncertainty of the result at small values of $\mu_\rmi{B}$ signifies the breakdown of the weak coupling expansion.

Finally, it should be noted that we have used the leading-order $m_\rmi{E}^2$ in generating both figs.~2 and 3, implying that in the high-$T$ limit there is a relative ${\mathcal O}(g^5)$ error in the results. This would be simple to correct by including the  ${\mathcal O}(g^4)$ correction to $m_\rmi{E}^2$ at high $T$.

\section{Discussion}
It is well-known that small but nonvanishing temperatures pose a technical problem for weak coupling expansions in dense quark matter. In this regime, it no longer suffices to treat only the static sector of the theory nonperturbatively, but the technical simplifications associated with the $T=0$ limit are not available either. While temperatures parametrically smaller than $m_\rmi{E}$ have been extensively studied \cite{Ipp:2003cj,Schafer:2004zf,Gerhold:2004tb,Gerhold:2005uu}, a connection to temperatures of order $\mu_\rmi{B}$ has been established only on a proof-of-principle level \cite{Ipp:2006ij}, and no EoS amenable to phenomenological applications exists.

In the letter at hand, we have addressed the challenge of small temperatures by formulating a new framework for high-order weak coupling calculations in deconfined QCD matter. Making use of known effective descriptions for the static and soft non-static sectors, we have derived a semi-analytic expression for the EoS, valid up to and incuding order $g^5$ at all values of $T/\mu_\rmi{B}$. The fact that our approach utilizes the framework of dimensional reduction to account for the static sector was shown to lead to a smooth interpolation of the pressure between known state-of-the-art results at low and high temperatures, as well as to a rapid convergence with increasing $T$. The new result is in addition not restricted to beta equilibrium, but is a function of independent quark chemical potentials. 

At exactly zero temperature, the state-of-the-art perturbative EoS of quark matter \cite{Kurkela:2009gj} has been widely used to describe the ultradense matter found inside neutron stars. The present work generalizes this result to nonzero temperature, which should lead to a reduction in the uncertainty of the EoSs used to model neutron star mergers. One concrete possibility to achieve this is to follow the strategy of \cite{Kurkela:2014vha} in deriving constraints for the behavior of the EoS at moderate density by requiring that it approaches the perturbative quark matter limit at high densities. We shall, however, leave such applications of our result, as well as its obvious extensions to nonzero quark masses \cite{Fraga:2004gz} and more economical parameterizations \cite{Fraga:2013qra}, for future work.

\section*{Acknowledgments}

We thank Eduardo Fraga, Ioan Ghisoiu, Tyler Gorda, Paul Romatschke, and Kari Rummukainen for useful discussions. The work of AV has been supported by the Academy of Finland, grant no.~273545.

\appendix

\begin{widetext}

\vspace{-0.2cm}

\section{Notation \label{notation}}

We work with $N_f$ flavors of massless quarks, keeping also the number of colors $N_c$ a free parameter. Various group theory factors defined using the generators of the fundamental representation of the gauge group SU($N_c$), $T^a$, as well as the structure constants $f^{abc}$ read
\ba
C_A \delta^{cd} & \equiv & f^{abc}f^{abd} \;=\; N_c \delta^{cd}, \\
C_F \delta_{ij} & \equiv & (T^a T^a)_{ij} \;=\; \fr{N_c^2-1}{2N_c}\delta_{ij},\\
T_F \delta^{ab} & \equiv & \tr T^a T^b \;=\; \fr{N_f}{2} \delta^{ab}, \\
d_A &\equiv& \delta^{aa} \;=\;  N_c^2 - 1, \\
d_F &\equiv& \delta_{ii} \;=\; d_A T_F/C_F \;=\; N_c N_f.
\ea

Independent chemical potentials $\mu_f$ are introduced for each quark flavor. In beta equilibrium they all agree, being related to the baryon chemical potential $\mu_\rmi{B}$ through $\mu_f=\mu_\rmi{B}/3$. We also introduce the shorthands 
\be
\mubar_f \equiv  \fr{\mu_f}{2\pi T}, \quad
z_f \equiv 1/2-\imathb\mubar_f
\ee
for variables that occur frequently in the results listed below, and in addition follow \cite{avpres} in defining the special functions
\ba
\zeta'(x,y) &\equiv& \partial_x \zeta(x,y), \label{specf1}\\
\aleph(n,z) &\equiv& \zeta'(-n,z)+\(-1\)^{n+1}\zeta'(-n,z^{*}), \\
\aleph(z) &\equiv& \fr{\Gamma '(z)}{\Gamma(z)}+\fr{\Gamma '(z^*)}{\Gamma(z^*)}, \ea 
where $\zeta$ is the generalized Riemann zeta function.




\section{Naive QCD pressure \label{pnaive}}

The $p_\rmi{QCD}^\rmi{naive}$ term appearing in our result has the form \cite{avpres}
\ba
\label{pe}
p_\rmi{QCD}^\rmi{naive} & = & p_1
 + g^2 p_2  + g^4
 p_3+ {\mathcal O}(g^6), \label{pqcdnaive}
\ea
where the functions $p_n$ read
\ba
p_1 &=& \fr{\pi^2}{45}\fr{T^4}{N_f}\!\sum_f\bigg\{d_A+\bigg(\fr{7}{4} + 30\mubar^2 + 60\mubar^4\bigg)d_F\bigg\},
\label{ae1} \\
p_2 &=& -\fr{d_A}{144}\fr{T^4}{N_f}\!\sum_f\bigg\{C_A + \fr{T_F}{2}\(1+12\mubar^2\)\(5+12\mubar^2\)\!\bigg\},
\label{ae2} \ea
\ba
p_3 &=& \fr{d_AT^4}{144(4\pi)^2}\Bigg[\fr{1}{N_f}\!\sum_f\bigg\{C_A^2\bigg(\fr{12}{\e}
+\fr{194}{3}\ln\fr{\bar{\Lambda}}{4\pi T} + \fr{116}{5} + 4\gamma -\fr{38}{3}\fr{\zeta'(-3)}{\zeta(-3)} +
\fr{220}{3}\fr{\zeta'(-1)}{\zeta(-1)}\bigg) \nn
&+& C_A T_F\bigg( \!12\(1+12\mubar^2\)\fr{1}{\e} +
\bigg(\fr{169}{3}+600\mubar^2-528\mubar^4\bigg)\ln\fr{\bar{\Lambda}}{4\pi T} +\fr{1121}{60} + 8\gamma \nn
&+& 2\(127+48\gamma\)\mubar^2 - 644\mubar^4
+ \fr{268}{15}\fr{\zeta '(-3)}{\zeta(-3)} + \fr{4}{3}\(11+156\mubar^2\)\fr{\zeta'(-1)}{\zeta(-1)} \nn
&+& 24\Big[52\,\aleph(3,z)
+ 144\imathb\mubar\,\aleph(2,z)+\(17-92\mubar^2\)\aleph(1,z)+4\imathb\mubar\,\aleph(0,z)\Big]\bigg) \nn
&+& C_F T_F \bigg(\fr{3}{4}\(1+4\mubar^2\)\(35+332\mubar^2\)-24\(1-12\mubar^2\)\fr{\zeta'(-1)}{\zeta(-1)} \nn
&-& 144\Big[12\imathb\mubar\,\aleph(2,z)-2\(1+8\mubar^2\)\aleph(1,z)
-\imathb\mubar\(1+4\mubar^2\)\aleph(0,z)\Big]\bigg) \nn
&+& T_F^2 \bigg(\fr{4}{3}\(1+12\mubar^2\)\(5 + 12\mubar^2\)\ln\fr{\bar{\Lambda}}{4\pi T}
+ \fr{1}{3}+4\gamma + 8\(7+12\gamma\)\mubar^2 + 112\mubar^4
- \fr{64}{15}\fr{\zeta'(-3)}{\zeta(-3)} \nn
&-& \fr{32}{3}\(1+12\mubar^2\)\fr{\zeta'(-1)}{\zeta(-1)}
- 96\Big[8\,\aleph(3,z) + 12\imathb\mubar\,\aleph(2,z) - 2\(1+2\mubar^2\)\aleph(1,z)
- \imathb\mubar\,\aleph(0,z)\Big]\bigg)\bigg\} \nn
&+& 288\,T_F^2\fr{1}{N_f^2}\sum_{f\,g}\bigg\{2\(1+\gamma\)\mubar_f^2\mubar_g^2
-\Big[\aleph(3,z_f+z_g)+\aleph(3,z_f+z_g^*) \nn
&+&4\imathb\mubar_f\Big(\aleph(2,z_f+z_g) + \aleph(2,z_f+z_g^*)\Big)
- 4\mubar_g^2\,\aleph(1,z_f) -\(\mubar_f+\mubar_g\)^2\aleph(1,z_f+z_g) \nn
&-&\(\mubar_f-\mubar_g\)^2\aleph(1,z_f+z_g^*)
-4\imathb\mubar_f\mubar_g^2\,\aleph(0,z_f) \Big]\bigg\}\Bigg], \label{ae3}
\ea
and the special functions are as defined above. The sums over $f$ and $g$ appearing here are taken over all $N_f$ quark flavors, and the gauge coupling $g$ is the renormalized one.

Using results from \cite{avthesis}, it can be shown that in the $T\to 0$ limit the above result reduces to the expressions
\ba
p_1 &=& \fr{C_A}{12 \pi^2}\!\sum_f\Big(\mu_f^4+2\pi^2\mu_f^2 T^2\Big)+{\mathcal O}(T^4), \label{p1smallT} \\
p_2&=& -\fr{d_A}{64\pi^4}\!\sum_f\Big(\mu_f^4+2\pi^2\mu_f^2 T^2\Big)+{\mathcal O}(T^4),\\
p_3 &=& \fr{d_A}{72(2\pi)^6}\sum_f\mu_f^4\Bigg\{-C_A\bigg(33\,\ln\fr{\bar{\Lambda}}{2\mu_f}+71\bigg)+\frac{153C_F}{4}
+N_f\bigg(6\,\ln\fr{\bar{\Lambda}}{2\mu_f}+11\bigg)-24\,\ln\, 2 \Bigg\}\nn
&-&\frac{d_A}{72(2\pi)^6}(11-12\gamma)\bigg(\sum_f \mu_f^2\bigg)^2 +\frac{d_A}{4(2\pi)^6}\sum_f \mu_f^2\sum_g \mu_g^2 \,\ln \,\frac{\mu_g}{2\pi T} \nn
&-&\frac{d_A}{24(2\pi)^6}\sum_{f>g}\Bigg\{(\mu_f-\mu_g)^4\ln\frac{|\mu_f^2-\mu_g^2|}{\mu_f \mu_g}+4\mu_f \mu_g(\mu_f^2+\mu_g^2)\ln\frac{\mu_f^2+\mu_g^2}{\mu_f\mu_g}
-(\mu_f^4-\mu_g^4)\ln\frac{\mu_f}{\mu_g}\Bigg\} \nn
&+& \fr{d_A T^2}{36(4\pi)^4}\sum_f\mu_f^2\Bigg\{C_A\bigg(\frac{72}{\epsilon}+300\,\ln\fr{\bar{\Lambda}}{2\mu_f}+152\,\ln\,\mubar_f+133+48\gamma+104\frac{\zeta'(-1)}{\zeta(-1)}\bigg)\nn
&+&C_F\bigg(48\,\ln\,\mubar_f+105+144\frac{\zeta'(-1)}{\zeta(-1)}\bigg) \nn
&+&N_f\bigg(24\,\ln\fr{\bar{\Lambda}}{2\mu_f}-8\,\ln\,\mubar_f-10+24\gamma-32\frac{\zeta'(-1)}{\zeta(-1)}\bigg)\Bigg\}\nn 
&+&\frac{2d_A T^2}{3(4\pi)^4}\sum_{f>g}(\mu_f^2-\mu_g^2)\,\ln \,\frac{\mu_f}{\mu_g} \, +{\mathcal O}(T^4). \label{p3smallT}
\ea
We note in particular the divergence of $p_3$ in the zero temperature limit, visible in the latter term on the second line.

\section{HTL contribution}

In this section, we briefly comment on the numerical evaluation of the HTL integral of eq.~(\ref{pHTLres}). It is convenient to convert the sum over the Matsubara modes to an integral on the complex $\omega$ plane, whereby we arrive at the relatively compact expressions
\ba
\sumint^{'}_K  \log \left[ 1 + \frac{\Pi_\rmi{T}(K)}{K^2} \right] &=&2 \int_k \left[ T \log\left(\frac{1- e^{- \omega_\rmi{T}(k)/T}}{1-e^{-k/T}}\right)+ \frac{1}{2}\left( \omega_\rmi{T}(k)-k \right)\right]\nn 
&-& \int_k \int_0^k \frac{d\omega}{\pi} \phi_\rmi{T}(\omega,k) \left[ 2 n_B(\omega)+1\right],
\ea
\ba
\sumint^{'}_K  \log \left[ 1 + \frac{\Pi_\rmi{L}(K)}{P^2} \right] &=&2 \int_k \left[ T \log\left(\frac{1- e^{- \omega_\rmi{L}(k)/T}}{1-e^{-k/T}}\right)+ \frac{1}{2}\left( \omega_\rmi{L}(k)-k \right)-\frac{T}{2}\log(1+\frac{m_\rmi{E}^2}{k^2}) \right] \nn
&-& \int_k \int_0^k \frac{d\omega}{\pi} \phi_\rmi{L}(\omega,k) \left[ 2 n_B(\omega)+1\right], \label{PiLHTL}
\ea
reminiscent of the results of \cite{Andersen:1999sf}. The only difference to the calculation presented in this reference is that we have explicitly removed the zero mode contribution from our sum-integrals. 

In both of the above results, the first lines originate from quasiparticle poles and the second ones from branch cut contributions. In the former terms, the functions $\omega_\rmi{T/L}$ denote the transverse and longitudinal plasmon frequencies, satisfying
\begin{align}
-\omega^2 + k^2  + \Pi_\rmi{T/L}(\omega_\rmi{T/L}(k), k) = 0
\end{align}
and having the well-known large- and small-$k$ expansions
\ba
\!\!\!\!\!\!\!\!\!\!\omega_\rmi{T}(k)  &\stackrel[k\ll m_\rmi{E}]{}{\approx}& \frac{m_\rmi{E}}{\sqrt{3}}+\frac{3 \sqrt{3} k^2}{5 m_\rmi{E}}-\frac{27  \sqrt{3} }{35}  \frac{k^4}{m_\rmi{E}^3}, \label{eq:wt_largek} \\
\omega_\rmi{T}(k)&\stackrel[k\gg m_\rmi{E}]{}{\approx}& k+\frac{m_\rmi{E}^2}{4 k}+\left(3 - 2 \log\left[ \frac{ 8 k^2}{m_\rmi{E}^2}\right]\right)\frac{m_\rmi{E}^4 }{32 k^3}, \nonumber \\
\omega_\rmi{L}(k) &\stackrel[k\ll m_\rmi{E}]{}{\approx}& \frac{m_E}{\sqrt{3}}+\frac{3\sqrt{3}k^2}{10 m_E^2}, \\
\omega_\rmi{L}(k) &\stackrel[k\gg m_\rmi{E}]{}{\approx}& k.
\ea
The branch cut contributions are on the other hand due to a cut in the function $\Pi_\rmi{HTL}$ for $|\omega| < k$, which contributes to the final result through
\ba
\phi_\rmi{T/L}(\omega,k) &\equiv& - \im \log\bigg[1+ \frac{\Pi_\rmi{T/L}(\omega,k)}{-\omega^2+k^2}\bigg] .
\ea
\end{widetext}


\end{document}